\documentstyle[epsf,graphics,a4,12pt]{article}
\begin{document}
\baselineskip=18pt
\def\be{\begin{equation}}
\def\ee{\end{equation}}
\def\bear{\begin{eqnarray}}
\def\eear{\end{eqnarray}}
\def\E{{\rm e}}
\begin{center}
{\Large\bf Decay amplitudes in two-dimensional QCD}
\vskip 1.1cm
{\large \bf E. Abdalla$^a$ and 
R. Mohayaee$^b$}\\
\vskip 0.3cm
{\it Instituto de 
F\'\i sica-USP, C.P. 66.318, S. Paulo, Brazil,\\
\vskip 0.3cm
$^a$eabdalla@fma1.if.usp.br
\hskip 1mm $^b$roya@fma1.if.usp.br}
\end{center}
\abstract 

\noindent
Decay amplitudes for mesons in two-dimensional QCD are discussed. 
We show that in spite of an infinite number of conserved charges, particle 
production is not entirely suppressed. This phenomenon
is explained in terms of quantum corrections to the 
combined algebra of higher-conserved and spectrum-generating currents. 
We predict the qualitative form of particle production probabilities and 
verify that they are in agreement with numerical data. We also discuss 
four-dimensional self-dual Yang-Mills theory in the light of our results.

\vskip .3cm
\noindent

\vfill\eject

\section{Introduction}
\indent

Major advances in two-dimensional QCD have been made since the pioneering 
work of 't Hooft \cite{thooft}, who found the mesonic spectrum of the theory 
in the limit of a large number of colours. In subsequent works, 
an exact expression for the decay amplitudes of the mesons was found 
\cite{ccg, bd}. However, both the 't Hooft
equation and the decay amplitudes could only be solved  
numerically. The numerical solution of the theory has been enterprised by
many authors\cite{burk,horn-b-pauli}.
The numerical calculations are difficult to
converge for massless fermions. A strongly convergent algorithm
for very reliable numerical computations is also available in the
literature\cite{ss}. These new 
results open challenging ways
of testing various aspects of the theory, specifically those concerning 
decay amplitudes of the theory.
  
It has been suggested \cite{elcio2} that massless two-dimensional QCD is an 
integrable system. It is believed that integrability of a 
theory implies stability of its bound states. One could, therefore, expect 
vanishing decay amplitudes for the mesonic states of 
the theory. The numerical results indicate otherwise. 
The decay amplitudes, however, are small relative to those obtained 
for the massive theory.

The main purpose of this article is to resolve this apparent 
conflict by relating diverse properties of two-dimensional QCD, as well
as understand more about the dynamical structure of two dimensional
QCD. 
Specifically, we show that the 't Hooft sector and the integrable sector 
of the theory are decoupled, in the sense that the conservation
laws found before do not commute with the spectrum generating algebra due
to quantum corrections.
We also believe that the theory is not integrable in the
't Hooft sector, that is to say that the integrability properties are 
spoiled by quantum corrections. However, we think that a strong 
simplification occurs in the massless case as compared to the massive 
fermion theory, allowing us to draw special attention to such a difference
in terms of the conservation laws and the anomalies, which presumably
have a mild contribution semi-classically.

We review the known bosonisation procedure in Section 2 and show how 
the theory can be re-expressed in a fermionic language,
by using gauge-invariant chiral fermions. 
The final re-fermionised action provides a convenient 
starting point for the analysis of
the mesonic spectrum. The significant outcome of this procedure is that the 
Poisson brackets receive quantum 
corrections. The conserved Noether currents then satisfy the 
Kac-Moody algebra only on taking the contribution from the 
anomalous Poisson bracket into account.
In Section 3, we combine the algebra of these 
conserved currents with that of the fermionic bilinears, which have 
been shown to generate a $W_\infty$ algebra. We show 
that, except for the pion, the mesons generated by the above $W_\infty $ 
algebra are not eigenstates of the Sugawara operator. 
Conservation of the Sugawara operator indicates the existence of an infinite
number of conservation laws. If such conservation laws survive 
quantization and
physical states are eigenstates of the Sugawara operator then the 
dynamics of the theory is severely constrained. On the other hand, the 
breakdown of conservation laws by anomalous terms  permits the decay of 
higher states. This explains the 
small decay rate numerically observed in the massless theory. Based on these 
results, we 
make detailed predictions, in Section 4, about various features of the 
spectrum. 
In Section 5, we use the recent numerical results to verify our predictions 
and hence the accuracy of our calculational results.   

\section{Bosonisation vs Fermionisation}
\indent

We start with the lagrangian
\be
{\cal L} = -{1\over 4}F_{\mu\nu}F^{\mu\nu} +i\overline\psi\gamma^\mu
D_\mu\psi , 
\label{1lag}
\ee
which describes QCD with massless fermions. In two dimensions, the theory 
can be bosonised by computing the fermionic determinant 
which arises in performing the path integration. In the 
bosonisation procedure one first defines left and right components of 
the gauge field in terms of the matrix-valued fields $U$ and $V$, {\it i.e.}, 
\be
A_ + ={i\over e}U^{-1}\partial_+U \quad  {\rm and} \quad
A_- ={i\over e}V\partial_-   V^{-1}.
\label{2adef}
\ee
It is well-known that the Jacobian resulting from the 
above change of variables together with the effective action can be 
expressed in terms of the WZW functional \cite{witten}. That is to say, 
\be
{\rm det} i\not \!\! 
D \equiv \E^{iW[A]} = \int {\cal D}g \E^{iS_F[A,g]}.\label{3det}
\ee
The bosonic action $S_F[A,g]$ may be obtained by the 
repeated application of the Polyakov-Wiegmann identity \cite{polwig,elcio1} 
and is given by, \begin{eqnarray}
S_F[A,g]&=&\Gamma [g] +{1\over 4\pi}\int {\rm d}^2x \big[ e^2 A^2 -
e^2A_+gA_-g^{-1} \nonumber\\
&&\quad\quad -ieA_+g\partial_-g^{-1}-ieA_-g^{-1}\partial_+g\big]\nonumber\\
&=& \Gamma[UgV] -\Gamma[UV],\label{4gaugeaction}
\end{eqnarray} 
where $\Gamma$ is the WZW action functional.
Subsequent inclusion of the Fadeev-Popov ghost action $S[gh]$, 
replacement of the variables $UV$ by $\Sigma$ and $UgV$ by $\tilde g$--using 
systematically the
invariance of the Haar measure, and finally the introduction of a scalar 
field $E$, to disentangle the $F_{\mu\nu}F^{\mu\nu}$ interaction, yields 
the following form of the partition function \cite{elcio1,elcio2}:
\begin{eqnarray}
{\cal Z}&=&\int {\cal D} \tilde g \E^{i\Gamma[\tilde g]}
{\cal D}[gh]\E^{iS[gh]}{\cal D}\Sigma {\cal D}E\E^{-i(c_V +1)\Gamma [\Sigma]}
\nonumber\\
&\times& \exp{[-(c_V +1){{\rm tr}}\int {{\rm 
d}}^2x\partial_+E\Sigma\partial_- \Sigma^{-1}
-2 e^2(c_V+1)^2 {{\rm tr}}\int {{\rm d}}^2 x E^2]}.
\nonumber\\
& &\label{5partition}
\end{eqnarray}

From a further replacement of the variable $E$ by a field $\beta$, 
satisfying the relation $\partial_+E={i\over
4\pi}\beta^{-1}\partial_+\beta $, the factorised form of the
partition function, 
\be 
{\cal Z}={\cal Z}[g,1]{\cal
Z}[\tilde\Sigma,-(c_V+1)]{\cal Z}_{gh} {\cal Z}_\beta,\label{6factor} 
\ee
is obtained. In the above expression, ${\cal Z}[M,n] $ is the 
partition function of a WZW field $M$ with central 
charge $n$, ${\cal Z}_{gh}$ is the ghost contribution,
the non-trivial coupling-constant-dependent part of the partition 
function is
\be 
{\cal Z}_\beta \equiv \int {\cal D}\beta{\cal D}C_-
\E^{i\Gamma[\beta]+{\rm tr}\int {\rm d}^2x\big[ {1\over
2}(\partial_+C_-)^2 +ieC_-\beta^{-1}\partial_+\beta\big]},
\label{7betapart} 
\ee 
and the auxiliary field $C_-$ is introduced to make the action local.
\footnote{The factorised form of the partition function is
actually elusive, due to several BRST constraints remaining from the gauge
condition and change of variables. For the definition of vacuum
and physical states the reader is referred to
\cite{elcio3,rothe1,rothe2}.}.
The equations of motion are equivalent to the conservation 
law $\partial_+I_-=0$, where
\be
I_- (x)= 4\pi e^2\beta^{-1}\partial_+\beta -2\Pi_-^\prime
+i(4\pi)^2e^3 C_--(4\pi e)^2 \big[ C_-,\Pi_-\big]\, ,\label{8I}
\ee
$\Pi_-=\partial_+C_-$ 
and $\Pi_\beta= {1\over 4\pi}\partial_+\beta^{-1} +
ie C_-\beta^{-1}$ are canonical momenta conjugate to $C_-$ and $\beta$
respectively. 
Canonical commutation relations lead to a Kac-Moody algebra obeyed by the 
current $I_-$. As in the case of conformally invariant theories, the 
conservation law $\partial_+ 
I_-=0$ implies that there is an infinite number
of conserved charges, $Q^{(n)}=\int I_- (x^-) (x^-)^n dx^-$.
These act on asymptotic $\beta$ fields as multiplications by powers of the 
(negative) component of the momentum times right-SU(n) transformation, {\it
i.e.},
\be
Q^{(n)^{ij}} \vert \beta ^{kl}(p)\rangle\sim p_-^n\delta^{jl}U^{jn}\vert 
\beta ^{kn}\rangle.
\ee 
This puts a stringent requirement on particle scattering,  
forbidding particle production or decay
\footnote{A detailed discussion of the action of conserved currents on 
$\beta$ field, and the consequences for $\beta$-scattering have been 
discussed in ref. \cite{elcio4}, where an exact S-matrix has been 
conjectured.}.

Recently, there have been suggestions \cite{ss} that the stability
of the spectrum can be 
tested in the framework of 't Hooft's mesons decay amplitudes. The
numerical computations of the amplitudes 
provide a detailed check of the claims presented here, namely that  
the infinite number of conservation laws does not imply absence of
particle production or stability of the meson and zero 
decay amplitude.

As mesons are fermionic bound states, the relevance of the above 
conserved quantities to the decay of mesons can be conveniently 
studied by 
writing the $\beta$ sector of the theory, defined in (\ref{7betapart}), 
in terms of fermions.
In the fermionisation procedure, the WZW term is equivalent 
to an action for massless fermions and the interactions are treated in 
perturbation 
theory by using the adiabatic principle of form invariance \cite {elcio5}.

Using these techniques, we obtain the following fermionised form of the 
action: 
\be
S=\int{\rm d}^2x \big[ \psi_+^{\dagger i}i (\delta^{ji}\partial_--
ieC_-^{ij})\psi_+^j +{1\over 2}{\rm tr} (\partial_+C_-)^2 +i 
{\psi^\dagger}^i_-\partial_+\psi^i_-\big].\label{9sfer}
\ee

The 't Hooft spectrum can be algebraically generated by using the 
spectrum-generating algebra of gauge-invariant 
fermion bilinears. Since the $\beta$ fields in (\ref{7betapart}),
as well as their fermionic replacements in 
(\ref{9sfer}), are gauge-invariant objects, the above action represents a 
chiral theory which has to be quantised using anomalous Poisson 
brackets \cite{rothe3} (see also chapters 13 and 14 of \cite{elcio5}). 
Indeed, the left-moving Noether currents,
\be
I_-^f =\psi^\dagger \psi -{2\over e} \Pi_-^\prime -{e\over 4\pi}
C_- + i\big[ C_-,\Pi_-\big],\label{10ifer}
\ee
generate a Kac-Moody algebra only if the anomalous Poisson Bracket 
(APB), 
\be 
\big\{ \Pi_-(x),\psi^\dagger \psi(y)\big\} _{{\rm APB}}= 
{e\over 4\pi}\delta(x-y),\label{11apb} 
\ee
is used \footnote{The bosonic formulation is 
advantageous since it contains information of order $\hbar$ already 
at the classical level.}.

The operator (\ref{10ifer}) (or (\ref{8I})) contains colour 
indices. However, in 't Hooft's formulation
one deals exclusively with colourless states and, therefore, it is
natural to consider the Sugawara operators 
\be
L(x) = : I_{ij}(x)I_{ji}(x):\quad ,\label{12leqi2}
\ee
where the colour indices are contracted and the divergent 
terms in the Wilson expansion (corresponding to the Kac-Moody algebra) 
are subtracted by means of a normal-ordering prescription.

\section{Classification of the mesonic states}

\indent

In this section, we review the action of the $W_\infty$ algebra on the 
spectrum. 
The spectrum-generating algebra, as obtained in \cite{specalg}, 
arises from the bilinears $M_{\alpha 
\gamma}(x,y)=\psi_\alpha^\dagger(x)\E^{ie\int_x^y 
A_\mu(\xi){\rm d}\xi^\mu}\psi_\gamma(y)$, where $\alpha$ and $\gamma$ 
are the chirality indices ($+$,$-$). In the massive case, these 
bilinears are related by the equations of motion. In the massless case, 
however, the mixed term, $M_{+-}$, corresponds to the 
integrable-$\beta$ sector, while the right-right term, $M_{--}$, 
corresponds to the usual meson bound states of 't Hooft. 
Therefore, we take
\be
M_{--}(x^-,y^-;x^+)=\psi^{i\dagger}_-(x^-,x^+)\psi^i_-(y^-;x^+)\label{14mmm}
\ee
as the spectrum-generating current.
In the momentum space of the $x^-$ variables, a classical solution
of the equations of motion obeying the Gauss constraint is
\be
\big(M_{--} (k_-,k_-^\prime;x^+)\big)_{\rm class}=\delta 
(k_--k_-^\prime)\theta(k_-). \label{15class}
\ee
Expanding $M_{--}$ around such a solution in the large $N$ limit 
\cite{specalg} and using
\be
M_{--}=\E^{{i\over\sqrt N}W}(M_{--})_{\rm class}\E^{-{i\over\sqrt 
N}W}\label{16m} 
\ee
one finds that the Fourier modes of $W$, which represent quantum 
fluctuation around classical solution, obey 
the 't Hooft equation. Thus, the $W_\infty$ algebra found for the bilinears
in \cite{specalg} is a spectrum-generating algebra.

We emphasise that in the massive theory, the full content of the theory is 
preserved in 
the individual chiral sectors, $M_{++}$ and $M_{--}$, which are related
to each other by the fermion equation of motion. In the massless case, 
however, we study the bilinears constructed from $\psi_+$. There 
is a mixing of the sectors and it does not suffice to study one 
sector on its own.

\section{The combined algebra}
\indent

In this section, we study the interplay between the 
$W_\infty$-spectrum-generating algebra and the 
conserved current (\ref{12leqi2}). We identify 
the bilinear $\psi^\dagger_+\psi_+$ as the pion and $\psi^\dagger_+ 
D^n\psi_+$ as 
the higher states obtained from the fermion bilinear $M_{++}$. 

The pion is an eigenstate of $L(x)$ \footnote{Recall that in 
the present fermionic 
formulation we have to use the anomalous Poisson bracket.}; 
\begin{eqnarray}
\big[ L(x), \psi^\dagger (y)\psi(y)\big] &=&{e^2\over 2(x-y)^2}\psi^\dagger
\psi(y) + {e^2\over 2( x-y)} \partial\big(\psi^\dagger \psi(y)\big) 
\nonumber\\
&& +{\rm \quad
regular \; terms}.\label{17lact1}
\end{eqnarray}
The higher-state fermion bilinears, however, due to various anomalous 
terms, are not eigenstates;
\begin{eqnarray}
&&\big[ L(x),\partial\psi^\dagger\psi(y)-\psi^\dagger (y)\partial\psi(y)\big] 
=\nonumber\\
&-&{e^2\over 4}\big[{2\over (x-y)^2}
\big(\partial\psi^\dagger\psi(y)-
\psi^\dagger \partial\psi(y) \big)\nonumber\\
&&- {1\over x-y} \partial
\big(\partial\psi^\dagger\psi(y)-\psi^\dagger\partial \psi(y)\big) 
-{2\over (x-y)^4}\big]\nonumber\\
&+& e^2\big[{1\over (x-y)^2}\big(\psi^{i\dagger}\psi^j\psi^{j\dagger}\psi^i
\big) -{1\over x-y}\big(\psi^{i\dagger}\psi^j\partial\big(
\psi^{j\dagger}\psi^i\big)\big)\big]\nonumber\\
&+&\big[ -{ie^2\over2}\big[\Pi^{kj}C^{ki}-C^{jk}\Pi^{ik}\big]
-16{e^3\over\pi}C^{ji}+e\partial_1\Pi^{ij}\big]\times\nonumber\\
&&\big[-{1\over x-y} 
\partial\big(\psi^{j\dagger}\psi^i\big) +{1\over (x-y)^2}\psi^{j\dagger}\psi^i
\big]\nonumber\\
& &+{\rm anomalous \; Poisson\; Brackets \; terms} +\; {\rm
regular\; terms}.\label{18lact2}
\end{eqnarray}

Using the above two equations, we can write down the general form of the 
algebra obtained from the 
action of the Sugawara operator on the physical states, that is,
\be
L\cdot M_n \approx {n+1\over (x-y)^2}M_n +{1\over x-y}\partial M_n+ {\rm
anomalous\; terms} .\label{lonmn}
\ee
Thus we see that, in the absence of 
anomalies, the n$^{th}$ state
is an eigenstate of $L$ with eigenvalue $n+1$. Taking this into 
account, we find that the action of $L$ on the in and out states is given by 
\begin{eqnarray}
(n-n_1-n_2-1)\langle M_{n_1} M_{n_2}\vert M_{n_j}\rangle&=&\int dx\langle 
\vert L_{out}-L_{in}\vert 
\rangle\nonumber\\
=\int dx \int_{-\infty}^\infty dt {d\over dt} \langle 
M_{n_1}M_{n_2}\vert
L\vert M_n\rangle   & =&{\rm anomalies},\label{loutminuslin}
\end{eqnarray}
which shows the existence of an infinite number of conservation 
equations. In the absence of the anomalous terms,
the decay amplitudes vanish. However, when anomalous terms are present
unphysical (non--mesonic) operators are 
introduced in the right-hand side. Therefore, the amplitude is 
non-vanishing. 
If one of the out states is a pion, then we will have the following set of 
recursive relations:
\begin{eqnarray}
(n-n_1-1)\langle M_0 M_{n_1}\vert M_n\rangle &=&\sum_X\langle M_0 X\vert 
M_n\rangle\nonumber\\
& &+\sum_Y\langle M_0 M_{n_1}\vert Y\rangle
\end{eqnarray}
where $X,Y$ are bound states of lower states and a pion state 
is always present on the right-hand side. The solution to the above 
relations is found by using 
\be
\langle M_0 M_0\vert Y\rangle=0,
\ee
which is obtained from the 1/N expansion of the decay amplitude and is 
valid up to second order. These solutions require the 
vanishing of the decays involving pions 
\footnote{
It is worth mentioning that the algebra is independent of the number of 
colours, and the anomalous term only makes a first-order contribution in 
the $1/N$ expansion.}.

From eqn. (\ref{18lact2}), we see that there are further corrections 
which come from both higher terms in the
Wilson expansion and anomalous Poisson bracket quantization of the
chiral fermions. These terms cannot cancel one
another, due to their different functional form (as an example, higher
$\Pi$ derivatives never arise from anomalous Poisson Brackets).  
There are further higher terms when one considers the 
gauge-covariantized current. 

The decay amplitudes in the massless theory (which is 
integrable in the absence of anomalies) are suppressed as 
compared to those of the massive 
theory (non-integrable). This is entirely due to the quantum corrections. 
Since the anomalous terms are of order $\hbar$, they 
disappear for quasi-classical decays. In the massive theory, in addition 
to quantum corrections, there are mass terms which spoil integrability. 
These terms do not vanish in the quasi-classical approximation. In the 
massless theory, this approximation is reliable 
for decays of highly-placed states ({\it e.g.} near the continuum limit 
) to large-momentum states. On the contrary, decays into small-momentum 
states, marked by interference terms, are highly influenced by  
quantum corrections which spoil quasi-classical approximation.

On the basis of the preceding results, we predict the following 
features:

1. The pion decouples in the massless fermion theory. Such a decoupling
is valid in all orders of the $1/N$ expansion.

2. For massless fermions, the decay of large-mass states into states with 
large momenta is severely suppressed.

3. The probabilities for further decays, although reduced, are not very 
small. For small $N$, the amplitudes are much smaller than the 
corresponding ones in the massive case, due to the enhanced significance 
of the larger-order terms in the $1/N$ expansion 
\footnote{We recall again that $1/N$ corrections are very 
important in the massive case which strengthens these predictions 
for groups such as $SU(2)$.}.

4. Finally, the decay amplitudes of very massive mesons vanish, 
unexceptionally, in the massless fermion case.

\section{Numerical back-up and large $N$ behaviour}

\indent

In this section, we verify these predictions by means of the numerical 
computation of the amplitudes (for predictions 1,2 and 3), as well as, by 
using large $N$ analysis (for prediction 4).

Amplitudes for meson decay were initially derived 
in the framework of $1/N$-expansion in \cite{ccg}, and          
in  more detail, including higher order corrections, in 
\cite{bd}.
The $1/N$ corrections vanish in the massless case \footnote{The 
lowest-order term, however, seems to give a non-vanishing contribution. 
This is discussed later in this section.}. This 
can be explained by studying the expression for the decay amplitude 
\cite{bd},
\begin{eqnarray}                                                              
{\cal A}&=&(1-{\cal C}) {1\over 1-\omega} \int_0^\omega {\rm d} 
x\phi_n(x)   
\phi_p({x\over\omega})\Phi_q({x-\omega\over 
1-\omega})\nonumber\\                
&&-(1-{\cal C}){1\over \omega}{\rm d} x \int_\omega^1 {\rm d}x 
\phi_n(x)       
\phi_q({x-\omega\over 
1-\omega})\Phi_p({x\over\omega})\nonumber\\                
&&+{1\over N}(1-{\cal C}){f_q\over 
1-\omega}\int_0^\omega                    
{\rm d}x 
\phi_n(x)\phi_p({x\over\omega}),\label{19ampl}                        
\end{eqnarray}                                                                
where $\omega ={k_+^p\over k_+^n}$, $\Phi_n(x)=\int_0^1 {\rm d}y 
{\phi_n(y) \over (x-y)^2}$ and ${\cal C}$ denotes the interchange of final 
states.   

We observe that the higher-order corrections are 
always multiplied by the factor            
\be                                                                             
f_n=\int_0^1 {\rm d}x 
\phi_n(x),\label{13fn}                                     
\ee                                          
where $ \phi_n(x)$ is 't Hooft's wave function of the decaying 
state. It can be verified using 't Hooft's wave equation, that $f_n$ 
vanishes for massless fermions \cite{ccg}. Moreover, the 
authors of ref. \cite{bd} claim that, as higher-order corrections amount 
to a redefinition of 
constants in the massless case, only the very first term in the $1/N$ 
series survives. Because this does not hold in the massive 
case, corrections to all order exist. We restrict our 
analysis to the large-$N$ limit where amplitude for 
both massless and massive fermions may be 
non-vanishing.                      

In order to fully understand the behaviour of the leading-order 
term ({\it i.e.} the first term in       
expression (\ref{19ampl})) we need to 
solve the 't Hooft equation numerically. The solution is then inserted back 
into (\ref{19ampl}) 
and the numerical integration is carried out. We are then 
in a position to compare the results for the massless 
and the massive cases.

The method which employs a MATHEMATICA program is unsatisfactory because of
its bad-convergence behaviour. The numerical results are 
unreliable although
they are compatible with the predicted vanishing decay 
rate in the massless 
case, and the non-vanishing decay rate in the 
massive case. Therefore, a more sophisticated method of numerical 
computation is needed 
\footnote{For a survey of these methods 
using orthogonal eigenfunctions expansion we refer 
the reader to \cite{bd,hpp,jafmen}.} .

A more elaborate method \cite{ss} confirms 
that particles do not decay into states which contain the 
pion whereas in the massive 
case, the decay amplitudes are 
shown to be significantly large (see Figure 1).
\begin{figure}[htb]
\begin{center}
\leavevmode
\begin{eqnarray}
\epsfxsize= 4truecm\rotatebox{-90}{\epsfbox{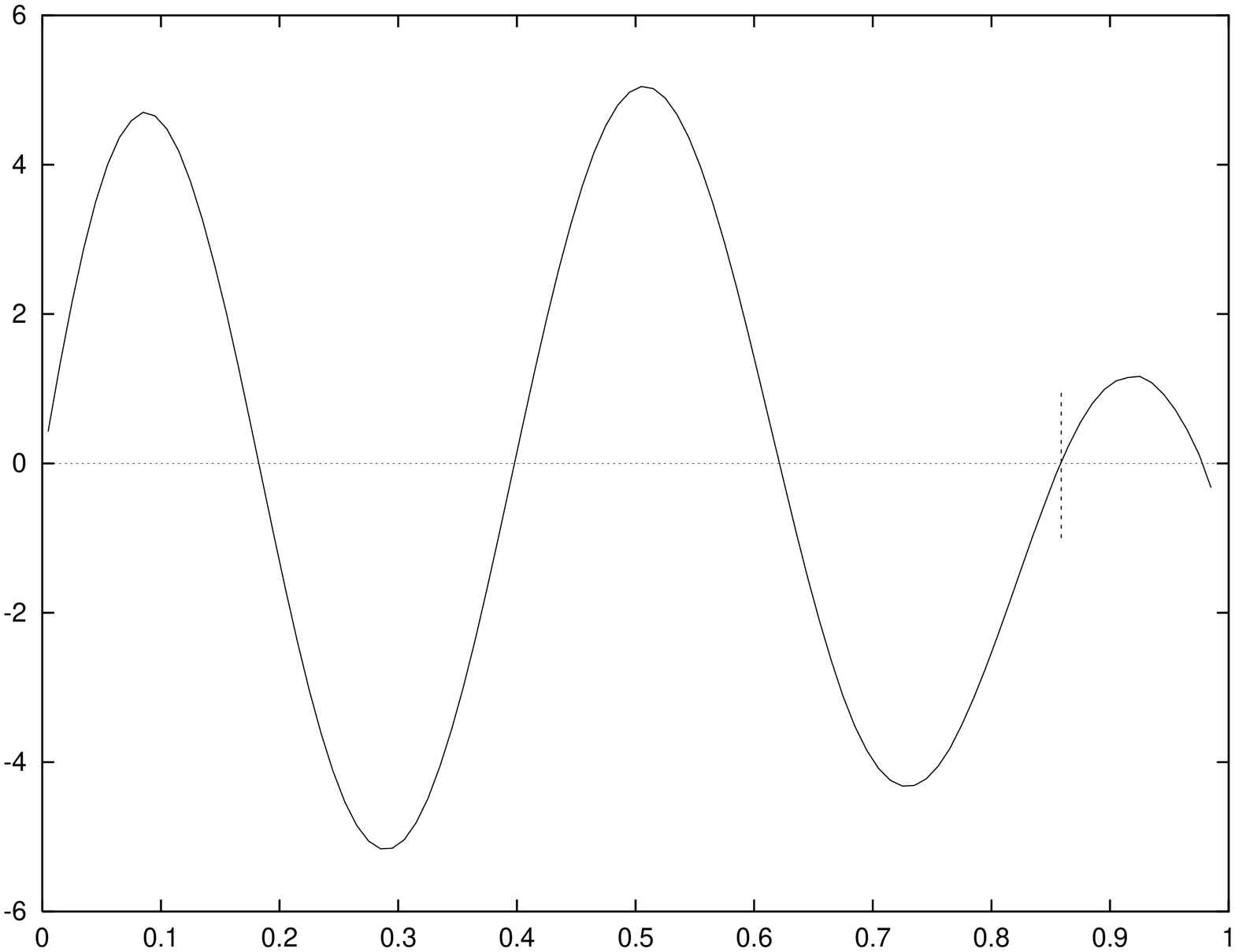}} & &
\epsfxsize=4truecm\rotatebox{-90}{\epsfbox{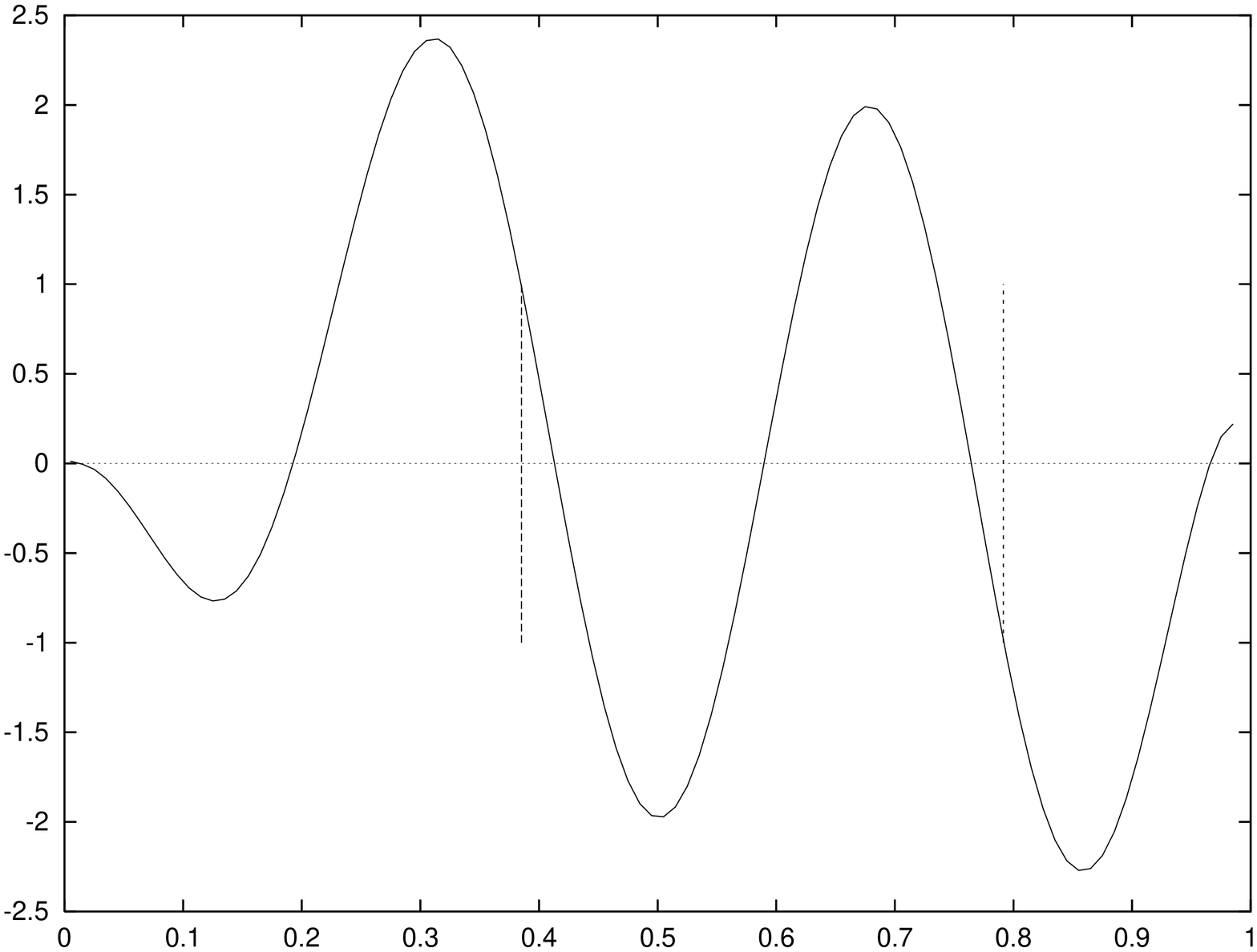}}\nonumber
\end{eqnarray}   
\vskip .5cm
\caption{{\it Decay rate of the $5^{th}$ excited state into the 
$1^{st}$ and 
ground (pion) states is plotted as a function of the outgoing momenta. The 
first diagram corresponds to massless and the second to massive 
fermions. The vertical bars mark the on-mass-shell values.}}
\end{center}
\end{figure}

For the massless case, several decays of the type $n\rightarrow 1+1$,
where 1 denotes the first-excited level of 't Hooft's series, have been 
computed (see Figure 3). For higher values of $n$ the amplitude approaches
zero rapidly. For the massive case, the 
amplitudes vary randomly (see Figure 4).

The overall results can be summarised as follows :

{\it (i).} In the massless case, for a general decay $k\rightarrow l+p$ the 
amplitude can start from a high value (see specifically the 
example $14\rightarrow 3+3$ in reference 
\cite{ss}) but decreases rapidly with increasing 
$k$ . These results are also compatible with the earlier results of 
reference \cite{bd}. Although the results for
the massless case are very precise, those for the massive case require 
further refinement. However, all the results obtained are precise enough for 
the conclusions drawn hitherto, and will be verified by other methods.

{\it (ii).} An exact computation of the decay rate of the higher-mass 
states guarantees
the correctness of the numerical results obtained so far. Indeed, using the 
large $N$ techniques of \cite{ccg} or \cite{michael} one finds
the required amplitude in the large $k$ limit. In terms of the 
fermion mass $m$, a parameter $\gamma$ satisfying
\be
\pi\gamma{\rm cot} \pi\gamma = 1-{\pi m^2\over e^2 N}\label{20mass}
\ee
is defined (the massless case corresponds to $\gamma=0$). In the large 
$n$ limit, the decay amplitude is found to obey 
\be
{\cal A}\sim \sin \pi\gamma ,\label{21largen}
\ee
which is valid up to constants describing the behaviour
of the 't Hooft wave functions at the origin (see \cite{michael} for 
further
details) and which shows a non-vanishing result only for non-zero values of 
the parameter $\gamma$. 
The same 
computation may be explicitly carried out in the massless case.

In order to investigate the issue of pair production for large number 
of colours and high mesonic states, we calculate the amplitude
of the decay of a very massive meson of mass $m_n$ into two
mesons of masses $m_{n_1}$ and $m_{n_2}$. Such an amplitude, at lowest 
order in
the inverse number of colours and up to an overall constant, is given by
the vertex
\bear
V_{{n_1}{n_2}n}&=&e^2{p_1\over p_2}\int_0^1 dx\int_0^1 dy {1\over 
\bigl(x{p_1\over p_2}\bigr)^2}\nonumber\\
&\times&\phi_{n_1}(x)\phi_{n_2}(y)\lbrack\phi_{n}({1-y\over 
1+{p_1/p_2}})-\phi_{n}({1+xp_1/p_2\over 1+p_1/p_2})\rbrack,\label{3vertex}
\eear
where $p_i$ is the plus--component of the momentum. For decay of very 
massive mesons, $m_n\to\infty$, energy-momentum conservation implies
\be
m^2_n=m_2^2{p_1\over p_2}.
\ee
Thus the main contribution to the integral comes from $x\sim 0$. Next, 
we make the following substitutions
\begin{eqnarray}
 x&=&\eta \bigl({m_2\over e}\bigr)^2,\\ 
\phi_i(x)&\approx& C_i x^\gamma,
\end{eqnarray}
 where $\gamma$ is defined in 
(\ref{20mass}) and $x$ is small, and 
\be
\phi_n \bigl(\xi\bigl({e\over m_n}\bigr)^2\bigr) \approx \phi (\xi)
\equiv \sin \bigl( {\xi\over \pi}+\delta (\xi)\bigr),\label{sin}
\ee
where $\delta (\xi)$ is a phase \footnote{The validity of  
equation (\ref{sin}) is confirmed for massless fermions by using 
numerical simulations.},in equation
(\ref{3vertex}) to find
\bear
V_{{n_1}{n_2}n}&=&\bigl({m_2\over m_n}\bigr)^{2\gamma}
C_1\int_0^\infty d\eta\int_0^1 dx' \eta^\gamma 
{\phi_2(x')\over (\eta+x')^2}\lbrack\phi((1-x'){m_2^2\over e^2})
-\phi((1+\eta){m_2^2\over e^2})\rbrack\nonumber\\
&=&\bigl({m_n\over e}\bigr)^{-2\gamma}C_1 I_2 .\label{initialintegral}
\eear
Since the vertex is symmetric under the exchange 
$1\leftrightarrow 2$,
the relation $C_i/ I_i=r$ must be the same for $i=1,2$ and can be 
computed in a convenient limit. We thus compute $r$ for $i=2$ and large 
values of $m_2$. In this limit, the asymptotic behaviour of the wave 
functions can be used to find the explicit values of $C_2$ and $I_2$, 
{\it i.e.},
\be 
C_2\approx \bigl({m_2\over e}\bigr)^{-2\gamma}\quad , \quad 
I_2\approx \bigl({m_2\over e}\bigr)^{2\gamma} \pi\sin \pi\gamma.\label{c2i2}
\ee
Subsequently, the expression
\be
V_{{n_1}{n_2}n}\approx e^2 \bigl({m_n\over e}\bigr)^{2\gamma}\sin\pi\gamma
\label{vertexofbeta}
\ee
is obtained which shows that the vertex vanishes for $\gamma\to 0$.

For massless fermions, that is for
$\gamma=0$,  the $\eta $ integration in the first term of eqn. 
(\ref{initialintegral}) 
can be performed. Moreover, by using 't Hooft's 
equation in the same term (to replace $\phi\over x$) we obtain
\begin{eqnarray}
\tilde I_2&=&-\int_0^1 dx\phi_2 (x)\int_0^\infty d\eta {\phi((1+\eta)\mu^2)
\over (\eta +x)^2} -\int_0^1 dx\phi_2 (x)\lbrack \mu^2+{1\over 
1-x}\rbrack\phi((1-x)\mu^2)\nonumber\\
&-&\int_0^\infty{d\xi\over \mu^2}\int_0^1 dx\phi(\xi) {\phi_2 (x)\over
(x-1+{\xi\over\mu^2})^2}
\end{eqnarray}
where $\mu={m_2\over m_n}$.
The last two terms cancel \cite{einhorn} due to the identity
\be
\int^\infty_0 d\zeta {\phi(\zeta)\over 
(\zeta-\eta)^2}=-({1\over\eta}+1)\phi(\eta).
\ee

In order to confirm these conclusions further, we make a detailed comparison 
between the amplitudes obtained in the massive and massless cases. 
This can be conveniently done by considering the table presented below. 
\vskip .3cm
\begin{tabular}{|c|c|c|c|}
\hline
 & & & \\  
{\it Decay series}&${\cal A}$& $k$ & ${\cal A}/k$\\
\hline
 & & & \\
$\quad 8\to1+1$, $m=0\quad$&$\quad$ .25 $\quad$ & $\quad$ .4 $\quad$ &  
$\quad$ .65  $\quad$    \\
\hline  
 & & & \\
$8\to1+1$, $m\not =0$ & .5       & .25 &   2         \\  
\hline 
& &  &\\ 
$9\to1+1$, $m=0$      & .2       & .4  &   .5        \\
\hline  
& & & \\ 
$9\to1+1$, $m\not =0$ &  .5      & .28 &   1.8      \\
\hline
& & & \\  
$10\to1+1$, $m\not =0$&  .8      & .3  &   2.7       \\
\hline  
& & & \\ 
$11\to1+1$, $m=0$     &  $\langle$.05    & .45 &   $\langle$.1       \\
\hline  
& & & \\ 
$13\to1+1$, $m=0$     &$\approx 0$& .45&$\approx 0$  \\
\hline   
\end{tabular}

\vskip .3cm
\noindent
In this table, the second column represents the momenta of the outgoing 
particles and the figures in the 
last column are proportional to the decay probabilities \footnote{We have 
chosen the units such that the mass of the initial 
state is unity (the influence of the fermion
mass is small in the present cases).}. These tabulated results once again 
confirm the prediction that the ratio between massless and
massive fermion decay rates goes to zero--being smaller than 3\% for 
the 11th state.
For other series of decays, this ratio approaches zero more slowly. 
Nevertheless, one clearly observes that this ratio approaches zero,
{\it e.g.}, for massless series $k\rightarrow 2+2$ (see F
igure 2).

\begin{figure}[htb]
\begin{center}
\leavevmode
\(\begin{array}{clr}
\epsfxsize=3truecm\rotatebox{-90}{\epsfbox{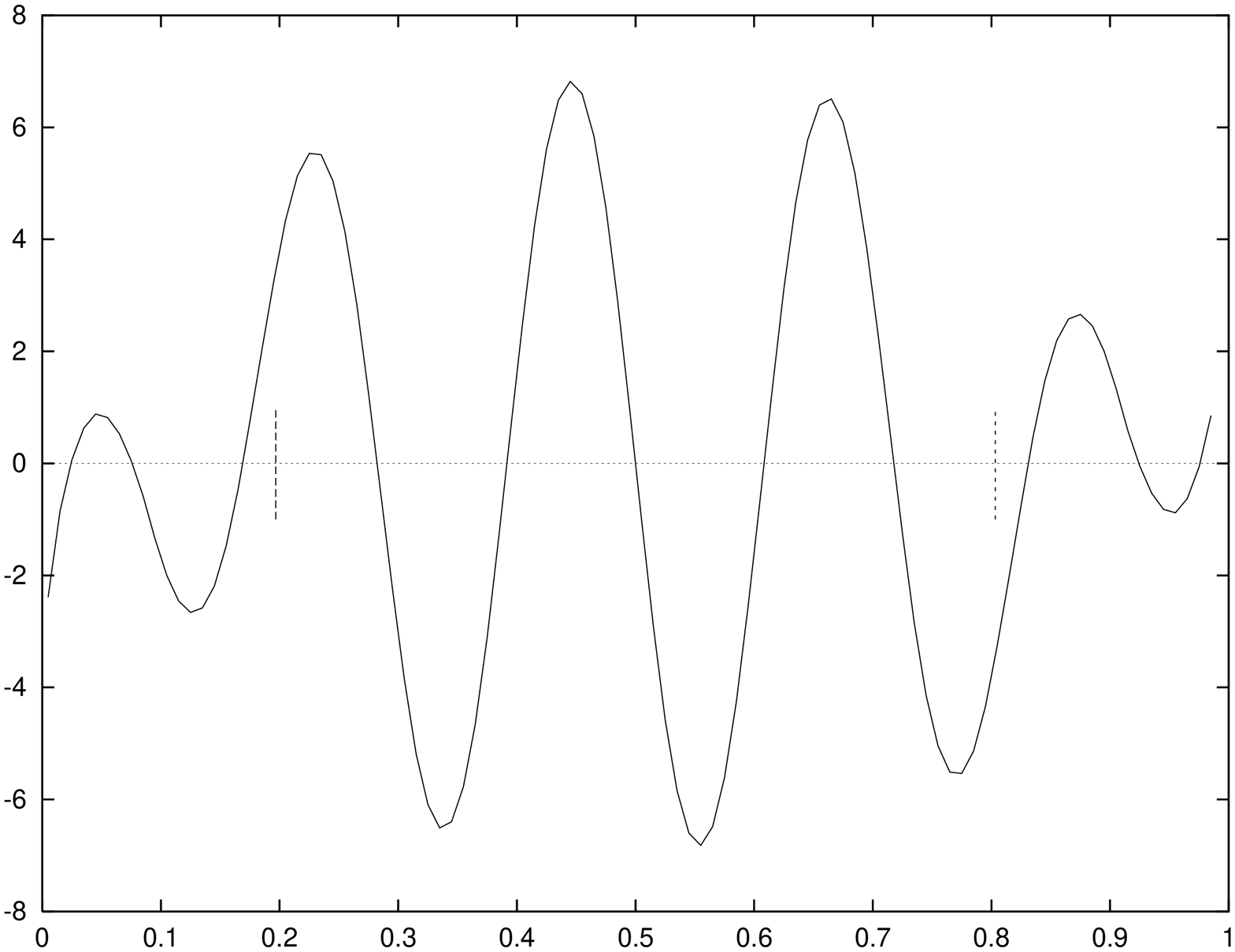}} & 
\epsfxsize=3truecm\rotatebox{-90}{\epsfbox{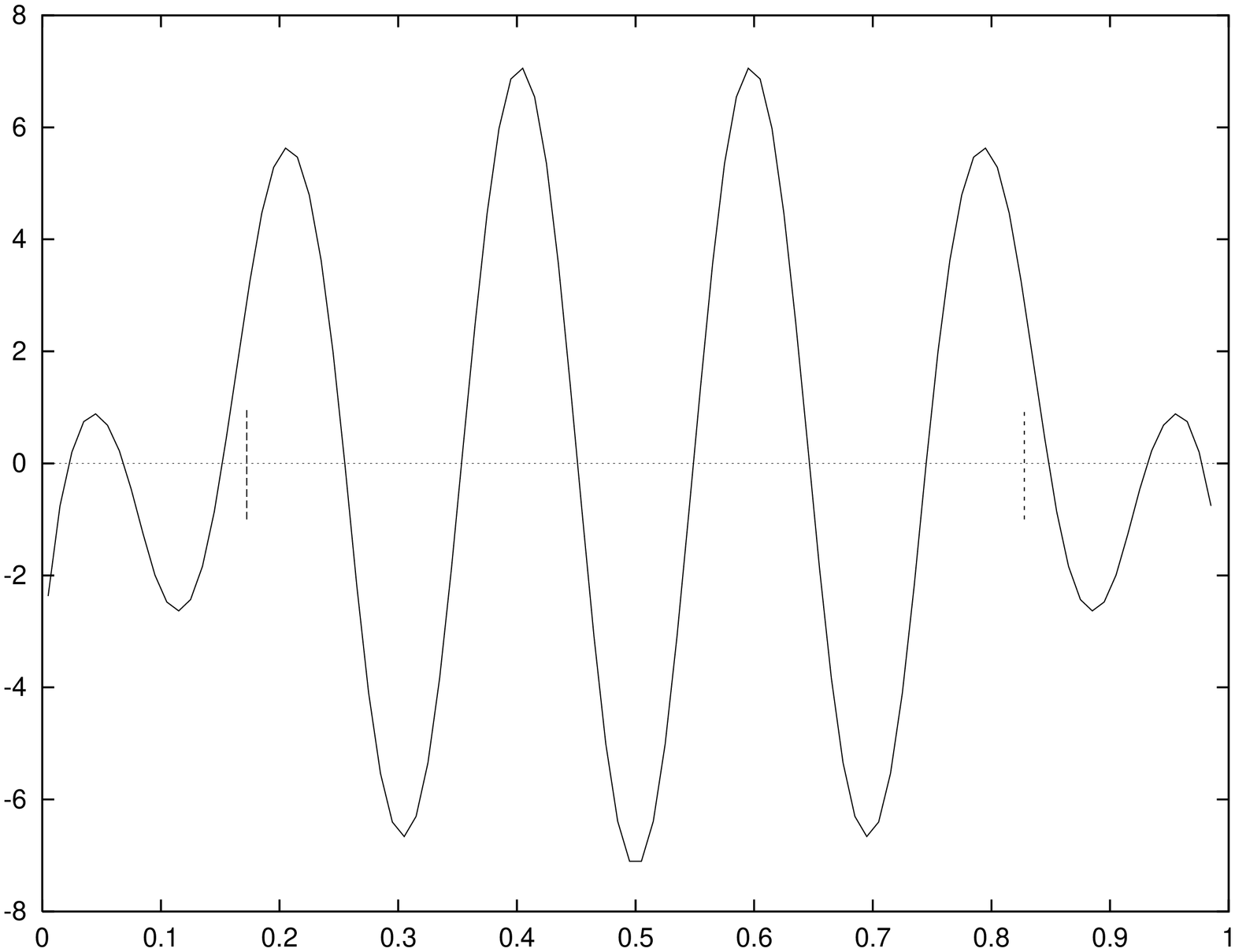}}&
\epsfxsize=3truecm\rotatebox{-90}{\epsfbox{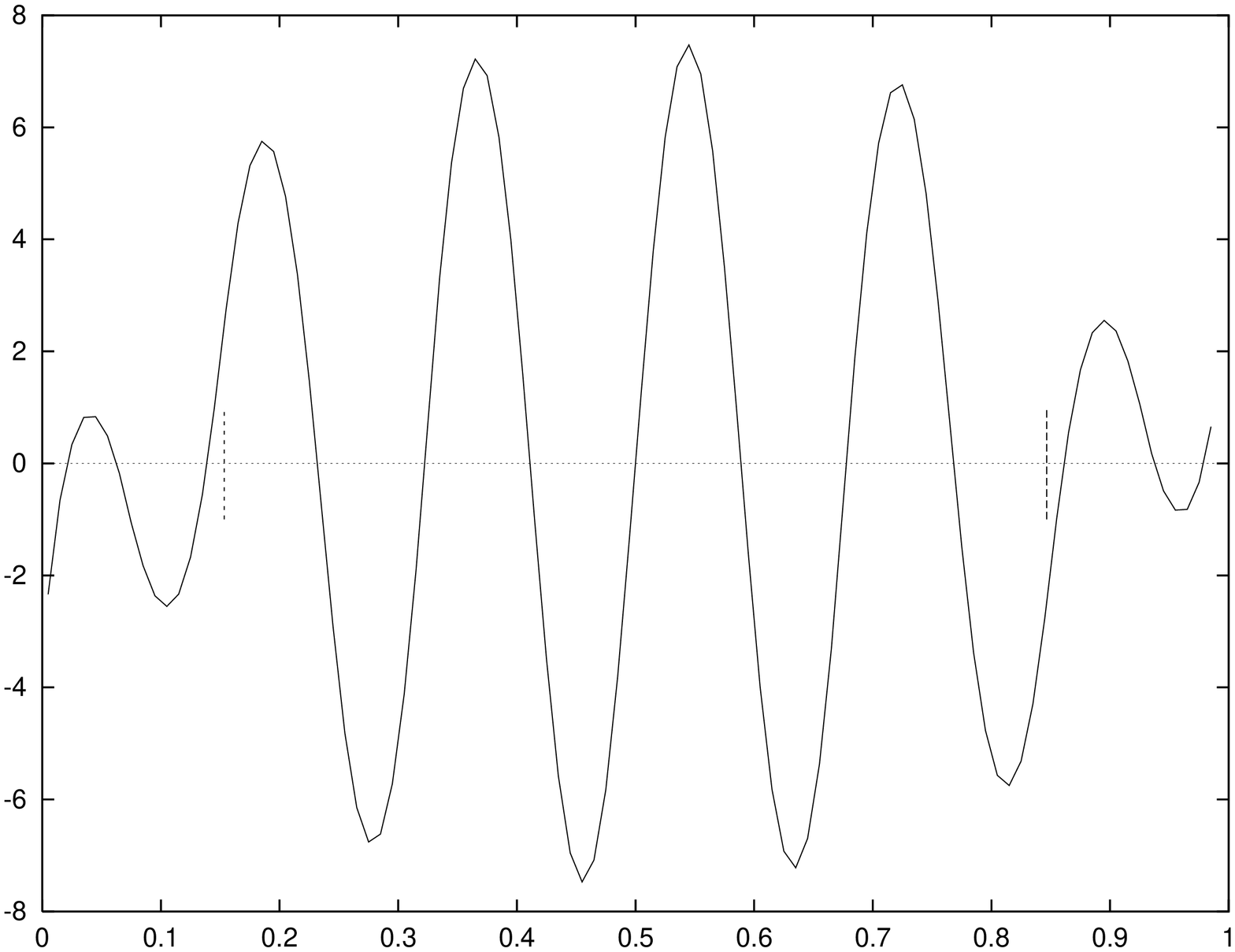}}
\nonumber
\end{array}  \) 
\vskip .5cm
 \caption{{\it Decay rates of the  
$10^{th}$,  $11^{th}$ and $12^{th}$ states into two mesons in the $2^{nd}$ 
excited state (for massless fermions) are plotted as functions 
of the outgoing momenta. The vertical bars mark the on-mass-shell values.}} 
\end{center}  
\end{figure}

\section{Conclusion}
\indent

Massless QCD contains higher-conservation laws which in general
imply integrability. These conservation laws have been derived 
in the massive ($\beta$) sector. The mesons in the 't Hooft sector are 
built up of fermion bilinears which are dressed with bosonic fields 
of the massless sector. We have shown here that the spectrum-generating
algebra, which defines the 
't Hooft sector, does not commute with the higher conservation 
laws due to quantum corrections to the short-distance expansions. 
This implies the 
break down of integrability in the meson sector. The quantum nature of
these corrections means that they are insignificant for 
quasi-classical decays. This renders the theory {\it quasi-integrable} 
and accounts for the exact decoupling of the pion. 

The theory has a complex structure of constraints.
The mesonic states, although 
physical, are not eigenstates of the conserved charges obtained from 
the Sugawara operators. This is because the massive and the massless 
sectors of the theory are connected by the constraint equations. The 
massive sector alone (the $\beta$ sector) is integrable but does not 
generate the physical Hilbert space since it is not BRST invariant. 
The change of variables made to decouple the dynamics of the 
massive and the massless sectors of the theory 
(see equations (\ref{2adef}), (\ref{5partition})
and (\ref{6factor})) leads to non-trivial jacobian which breaks the 
BRST invariance \cite{rothe1,rothe2}. 

Therefore, due to quantum corrections, the integrability and the 
BRST-invariance properties of the theory fall into two 
different sectors; the former 
being valid in the massive sector alone and the latter in the meson sector.
This can be interpreted as the stability of the unphysical $\beta$ particles.
We have verified this characteristic of the theory in the 
quasi-classical approximation by numerical methods.

The theory simplifies in the large $N$ limit, and meson wavefunctions 
and masses can be
computed. Large N corrections are well known in the literature, and the
numerical problem can be tackled.

Nevertheless, two dimensional QCD is far from trivial. In spite of the
large $N$ techniques, the formal aspects of the theory have not been
fully understood, and only recently the vacuum structure has been
studied, and separated from the description of the massive excitations.
Such structure are the core of the understanding of the Schwinger model,
and led to very profound consequences in that case. Our aim has
been to deepen the understanding of the theory, obtaining results
similar to those known to two dimensional QED, the Schwinger model.

However, it is clear that a development in that direction encounters a 
wall, since the theory is not soluble. Nevertheless, the numerical methods 
used, permit to obtain new structures otherwise
unknown. Moreover, there are indications that the theory has an unexpected
simplification in the massless case.

The result of our paper is to show that in spite of the complexity of the
situation, and the fact the the model is used to study QCD mesons, the
decaying amplitudes are simpler than imagined before, and indicate
further structures not known before. The zeroes of the decaying
amplitudes are a demonstration of that fact. The theory is not integrable,
but the decay amplitudes are nearly vanishing. Our numerical results
are an "experimental" constatation of that fact, and in section () we
give a field theoretic argument to support that fact.

Finally, we wish to point out that the methods are not borrowed from
techniques invented for integrable systems, but rather well defined and
established techniques based on the computation of the exact fermionic 
determinant, leading to the bosonised version, namely the gauge Wess Zumino 
Witten model and its gauge interaction.

It is of prime importance to generalise the concept of 
quasi-integrability to higher dimensions. Indeed, Bardeen 
\cite{bardeen} has recently pointed out that helicity amplitudes in 
high-energy QCD
are very simple at tree level and are described by a self-dual Yang-Mills 
theory. The classical solution of this theory strongly resembles the 
Bethe Ansatz solution of integrable two-dimensional models. Moreover, the 
one-loop amplitudes are reminiscent of those corresponding to 
anomalous conservation laws. It is known that the 
self-dual Yang-Mills theory 
is an integrable theory and is described by very simple actions 
\cite{chalmers}.
On the other hand, integrable Lagrangians with either anomalies 
\cite{elcio6} or with non-vanishing amplitudes for particle 
production\cite{cosmas}
are known and are well documented in the literature. It remains an 
interesting open problem to see whether the quasi-integrability idea is the 
most efficient framework for the description of non-trivial dynamics in 
theories with higher conservation laws, in general space-time 
dimensions, in spite of the Coleman-Mandula no-go theorem 
\cite{coleman} and its more general version \cite{haag}.

\section{ Acknowledgements}
\noindent

We wish to thank Chand Devchand for a critical 
reading of the manuscript and many useful suggestions.


\vskip 2cm

\begin{figure}[htb]
\begin{center}
\leavevmode
\begin{eqnarray}
\epsfxsize=3.5truecm\rotatebox{-90}{\epsfbox{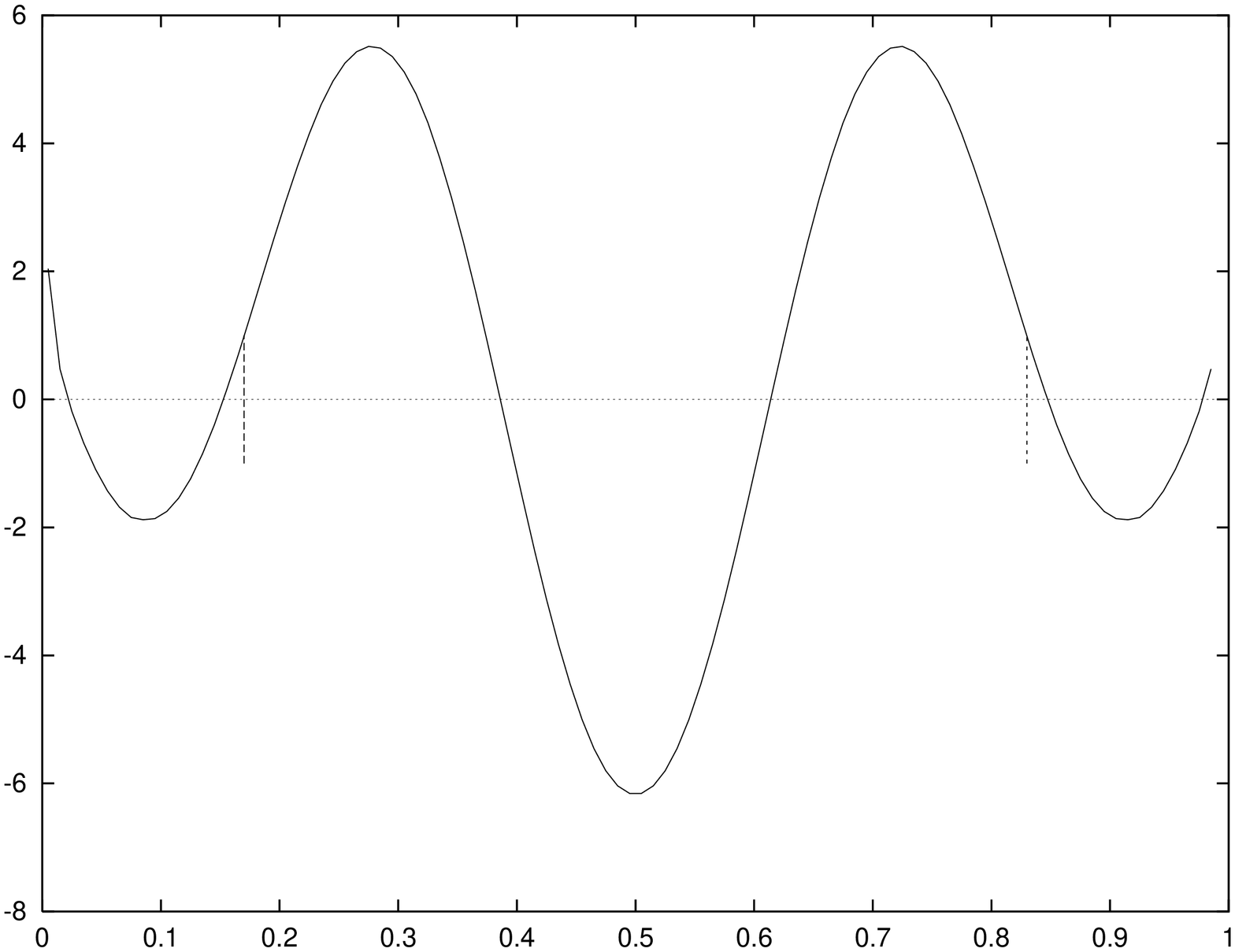}} & &
\epsfxsize=3.5truecm\rotatebox{-90}{\epsfbox{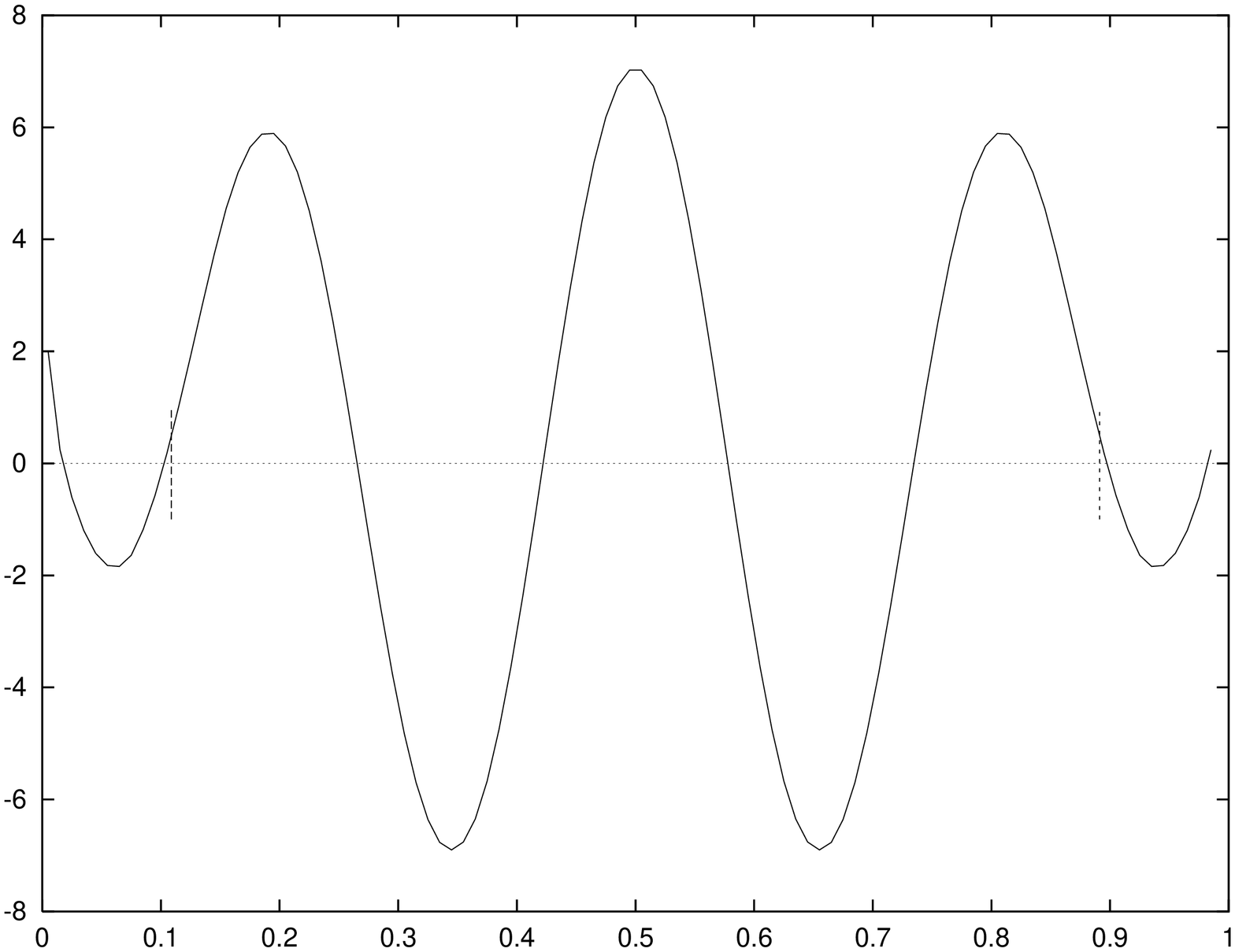}}\nonumber\\
\epsfxsize=3.5truecm\rotatebox{-90}{\epsfbox{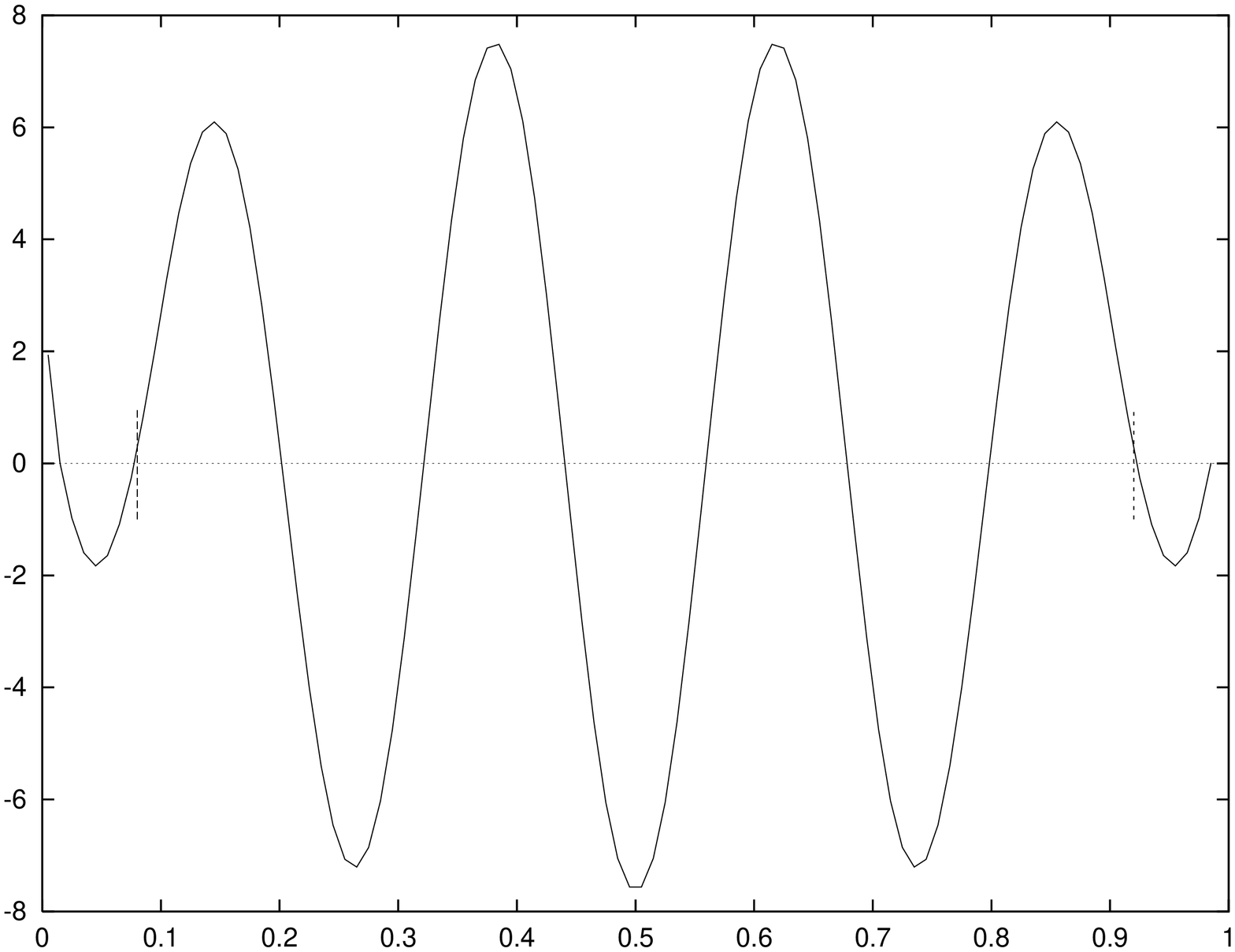}} & &
\epsfxsize=3.5truecm\rotatebox{-90}{\epsfbox{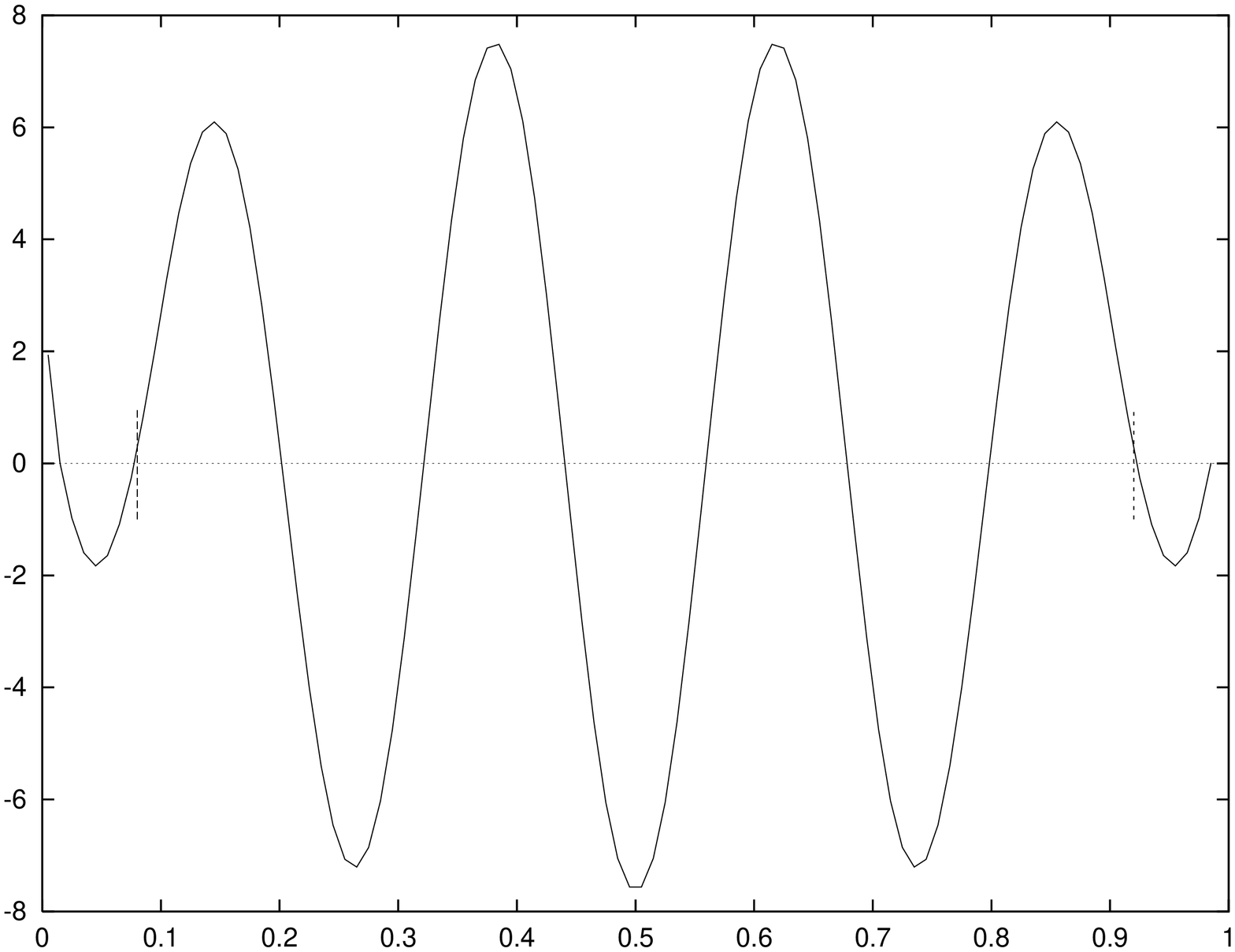}}\nonumber
\end{eqnarray}
\caption{{\it Decay rates of the $5^{th}$, $7^{th}$ , $9^{th}$ and $11^{th}$ 
excited state into two mesons in the 
$1^{st}$ excited state are plotted as functions of the outgoing momenta for massless fermions. 
The vertical bars mark the on-mass-shell values.}}
\begin{eqnarray}
\epsfxsize=3.5truecm\rotatebox{-90}{\epsfbox{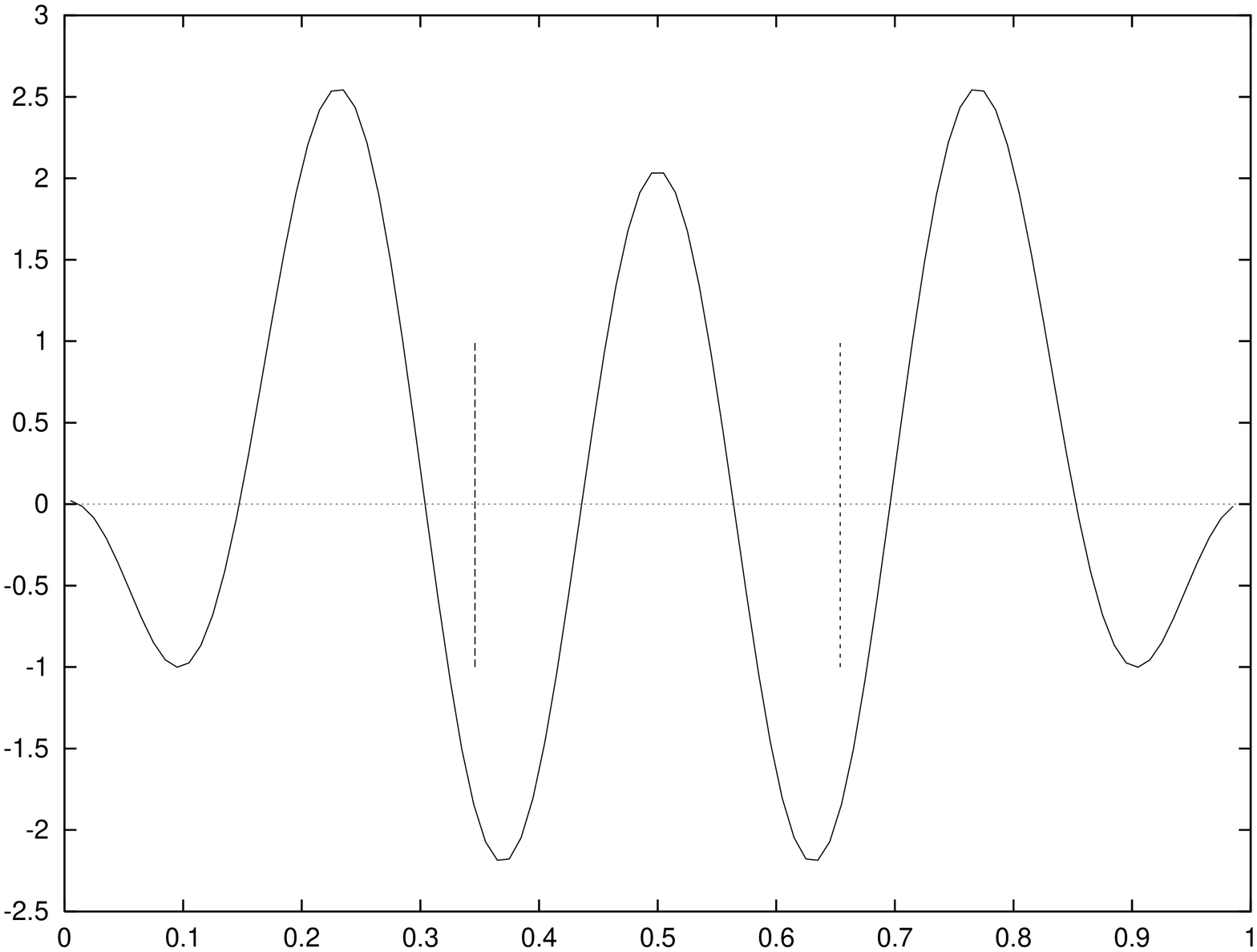}} & &
\epsfxsize=3.5truecm\rotatebox{-90}{\epsfbox{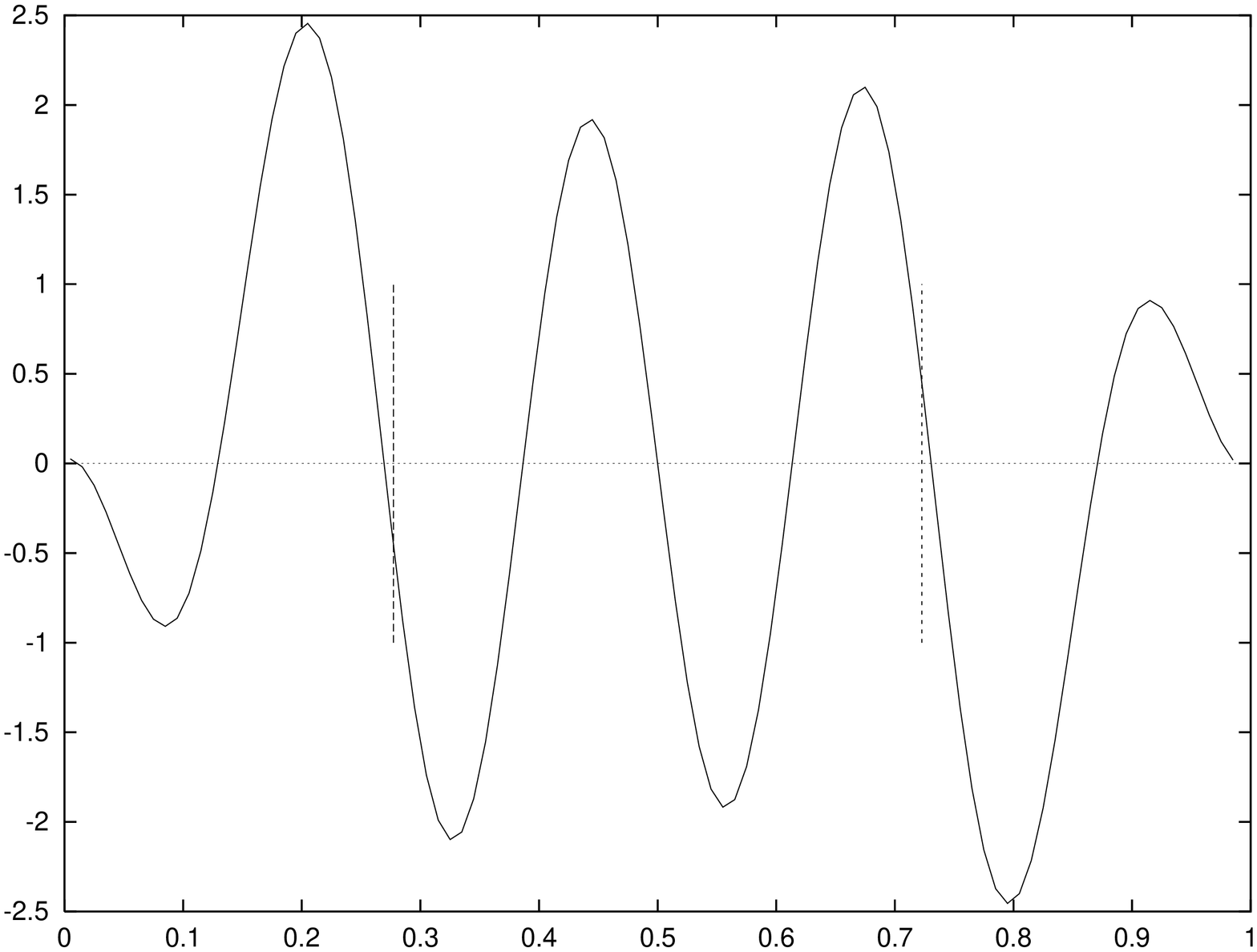}}\nonumber\\
\epsfxsize=3.5truecm\rotatebox{-90}{\epsfbox{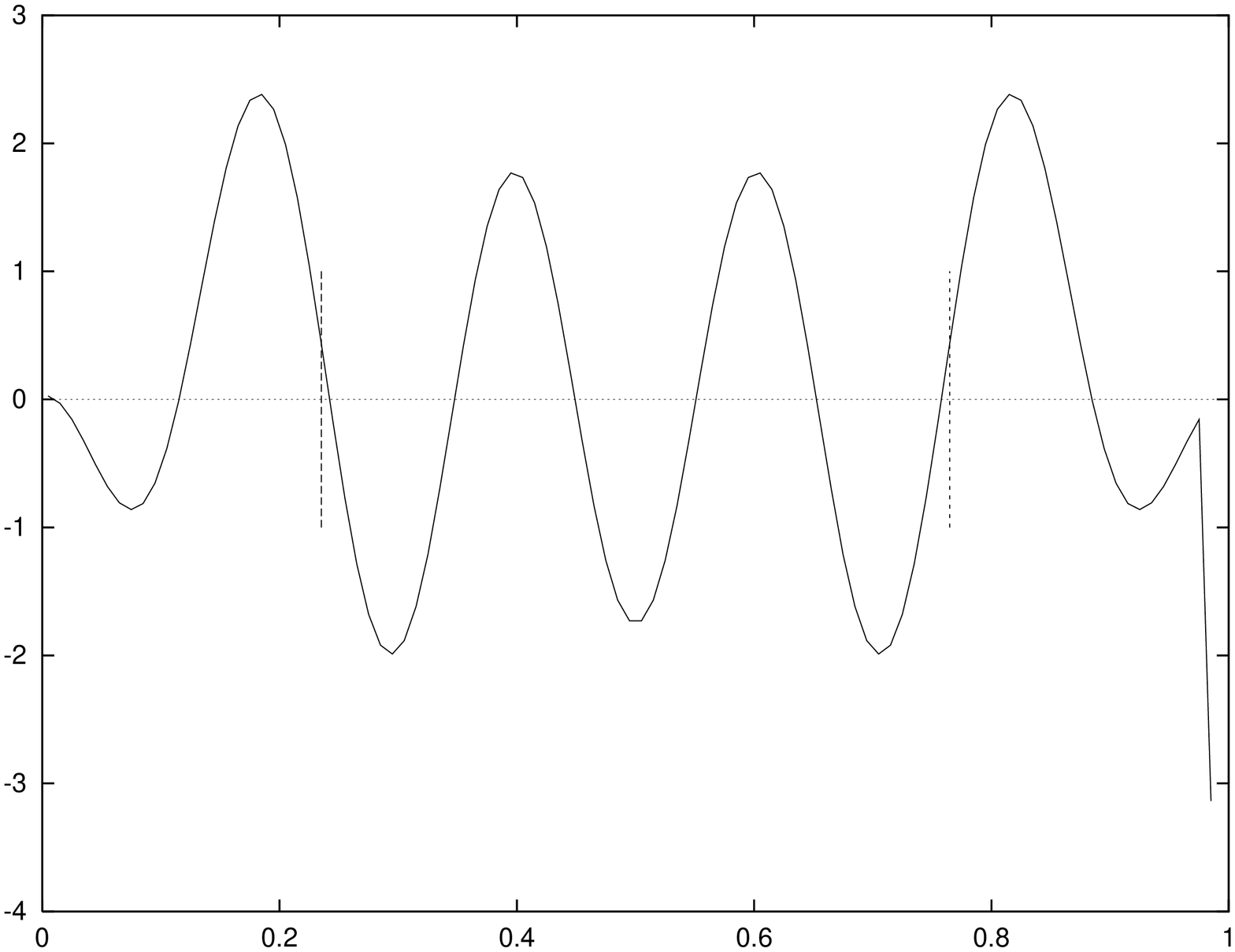}} & &
\epsfxsize=3.5truecm\rotatebox{-90}{\epsfbox{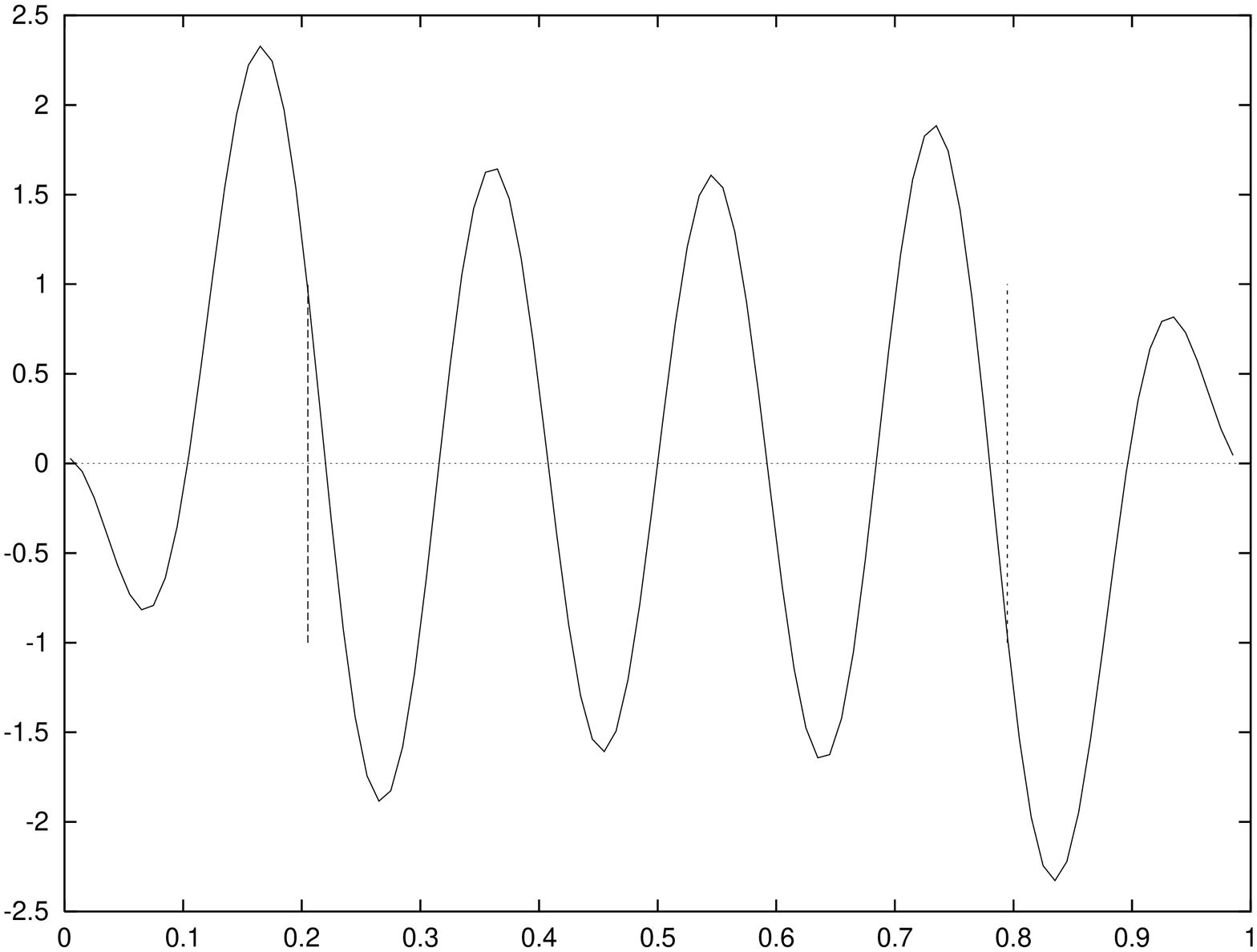}}\nonumber
\end{eqnarray}   
\caption{{\it Decay rates of the $7^{th}$, $8^{th}$, $9^{th}$ and $10^{th}$
states into two mesons in the $1^{st}$ excited state are plotted for 
massive fermions.}}
\end{center}
\end{figure}



\end{document}